\documentstyle[12pt]{article}

\def\thebibliography#1{\bigskip\section*{}\bigskip\list
{$^{\arabic{enumi}}$}{\settowidth\labelwidth{#1}\leftmargin\labelwidth
\advance\leftmargin\labelsep
\usecounter{enumi}}
\def\newblock{\hskip .11em plus .33em minus .07em}
\sloppy\clubpenalty4000\widowpenalty4000
\sfcode`\.=1000\relax}

\def\op#1{\mathop{\fam0 #1}\limits}

\newcommand{\Ker}{{\rm Ker\,}}

\newcommand{\di}{{\rm dim\,}}

\newcommand{\beq}{\begin{equation}}
\newcommand{\eeq}{\end{equation}}
\newcommand{\ben}{\begin{eqnarray}}
\newcommand{\een}{\end{eqnarray}}
\newcommand{\be}{\begin{eqnarray*}}
\newcommand{\ee}{\end{eqnarray*}}
\newcommand{\bea}{\begin{eqalph}}
\newcommand{\eea}{\end{eqalph}}
\newcommand{\bS}{{\bf S}}
\newcommand{\rA}{{\rm Ann\,}}
\newcommand{\cL}{{\cal L}}
\newcommand{\cV}{{\cal V}}
\newcommand{\cE}{{\cal E}}
\newcommand{\cH}{{\cal H}}
\newcommand{\bL}{{\bf L}}
\newcommand{\bR}{{\bf R}}

\newcommand{\bt}{\beta}
\newcommand{\dl}{\delta}
\newcommand{\la}{\lambda}

\newcommand{\g}{\gamma}
\newcommand{\G}{\Gamma}

\newcommand{\ve}{\varepsilon}
\newcommand{\th}{\theta}

\newcommand{\vt}{\vartheta}
\newcommand{\si}{\sigma}
\newcommand{\Si}{\Sigma}
\newcommand{\bom}{{\bf\Omega}}
\newcommand{\w}{\wedge}
\newcommand{\wt}{\widetilde}
\newcommand{\wh}{\widehat}
\newcommand{\ol}{\overline}
\newcommand{\dr}{\partial}

\newcommand{\ar}{\op\longrightarrow}
\newcommand{\ot}{\otimes}
\newcommand{\ap}{\approx}

\let\ssection=\section
\renewcommand{\section}{\setcounter{equation}{0}\ssection}

\newcounter{eqalph}
\newcounter{equationa}
\newcounter{theorem}
\newcounter{proposition}
\newcounter{lemma}
\newcounter{corollary}
\newcounter{definition}

\newenvironment{eqalph}{\stepcounter{equation}
\setcounter{equationa}{\value{equation}}
\setcounter{equation}{0}

\begin{eqnarray}}{\end{eqnarray}\setcounter{equation}{\value{equationa}}}

\def\thedefinition{\arabic{definition}}

\newenvironment{proof}{\noindent 
{\it Proof:}}{\medskip}
\newenvironment{rem}{\medskip\noindent{\it Remark:}}{\medskip}

\newenvironment{prop}{\refstepcounter{definition} 
\bigskip\noindent{\it Proposition \thedefinition:}}{\medskip}
\newenvironment{lem}{\refstepcounter{definition} 
\bigskip\noindent{\it Lemma \thedefinition:}}{\medskip}

\begin{document}
\hbox{}

{\parindent=0pt

{\large\bf Nonholonomic constraints in time-dependent mechanics}
\bigskip

{\sc G.Giachetta and L.Mangiarotti}\footnote{Electronic mail:
mangiaro@camserv.unicam.it}

{\sl Department of Mathematics and Physics, University of Camerino, 62032
Camerino (MC), Italy}
\medskip

{\sc G. Sardanashvily}\footnote{Electronic mail:
sard@grav.phys.msu.su}

{\sl Department of Theoretical Physics, 
Moscow State University, 117234 Moscow, Russia}
\bigskip
 
The
constraint reaction force of ideal nonholonomic constraints in time-dependent
mechanics on a configuration bundle $Q\to \bR$ is obtained. Using the
vertical extension of  Hamiltonian formalism to the vertical tangent bundle
$VQ$ of $Q\to\bR$, the Hamiltonian of a
nonholonomic constrained system is constructed.
}
\bigskip

\noindent 
{\bf I. INTRODUCTION}
\bigskip

This work addresses the geometric theory
of nonholonomic constraints in time-dependent mechanics. We refer the reader
to Refs. 1-7 for the 
autonomous case. We follow the approach based on the D'Alembert principle
because the variational methods with Lagrange multipliers are not
always appropriate to nonholonomic constraints (see Refs. 2,5,6,8).

Let the jet manifold $J^1Q$ be a velocity phase space of time-dependent
mechanics on a configuration bundle $Q\to\bR$. The most general 
nonholonomic constraints  considered in the literature are given by 
codistributions
$\bS$ or, accordingly, by  distributions $\rA(\bS)$ on the jet manifold
$J^1Q$.$^{9,10}$ 
Distributions on a configuration
space $Q$ and
submanifolds of the jet
manifold
$J^1Q$ can be also seen as nonholonomic constraints.$^{8,11,12}$

Dealing with nonholonomic constraints in time-dependent mechanics, one
usually studies the following problem. Let
$\xi$ be a second order dynamic equation on
$Q$ and
$\bS$ a codistribution on $J^1Q$ whose annihilator $\rA(\bS)$ is treated as a
nonholonomic constraint. 
The goal is to find a decomposition,
\beq
\xi=\wt\xi +r, \label{gm445}
\eeq
where $\wt\xi$ is a second order dynamic equation obeying the condition
\beq
\wt\xi\subset \rA(\bS). \label{gm446}
\eeq 
One can think of $\wt\xi$ as describing a mechanical
system subject to the nonholonomic constraint $\bS$, while $(-r)$ is 
the constraint reaction acceleration. The decomposition
(\ref{gm445}) however is not unique. In the case of Newtonian systems,
including nondegenerate Lagrangian systems, we obtain the decomposition
(\ref{gm445}) which satisfies the  D'Alembert principle for ideal
nonholonomic constraints. We construct the Hamiltonian
counterpart of the constrained equation of motion (\ref{gm446}). We show that
this can be seen as Hamilton equations in the framework of the
vertical extension of Hamiltonian formalism to the configuration space $VQ$
which is the vertical tangent bundle of $Q\to\bR$. This may be a step towards 
the functional integral formulation of nonholonomic time-dependent mechanics
and its further quantization.

\bigskip 

\noindent 
{\bf II. GEOMETRIC INTERLUDE}
\bigskip

All manifolds throughout the paper are real, finite-dimensional,
second-countable  (hence, paracompact) and connected.

We refer the reader to Refs. 8-11,13-15 for the geometric formulation of
Lagrangian and Hamiltonian time-dependent mechanics. In accordance with this
formulation, a configuration space of time-dependent mechanics
is  an $(m+1)$-dimensional fiber bundle $Q\to \bR$, coordinated by
$(t,q^i)$. Its base $\bR$ is treated as a time axis provided with the
Cartesian coordinate $t$.
With this coordinate,
$\bR$ is equipped with the standard vector field $\dr_t$ and the standard
1-form $dt$. For the sake of convenience, we will also utilize the compact
notation $q^\la$, where $q^0=t$. Obviously, any fiber bundle $Q\to\bR$
is trivial, but it cannot be canonically identified
to a product $\bR\times M$ in general. Different trivializations  $Q\cong
\bR\times M$ correspond to different reference frames.

The velocity phase space of time-dependent mechanics is the first order jet
manifold $J^1Q$ of $Q\to \bR$, coordinated by $(t,q^i,q^i_t)$. There is the
canonical imbedding,
\beq
 \la: J^1Q\hookrightarrow TQ, \qquad
 (t,q^i,q^i_t) \mapsto (t,q^i,\dot t=1, \dot q^i=q^i_t), \label{z260}
\eeq
of $J^1Q$ onto the affine sub-bundle of the tangent bundle $TQ$  of
$Q$ which is modelled over the vertical
tangent bundle $VQ$ of $Q\to\bR$. From now on
we will identify the jet manifold
$J^1Q$ with its image in
$TQ$. 

Similarly, we have the imbeddings,
\be
&& J^2Q\hookrightarrow J^1J^1Q\hookrightarrow TJ^1Q,\\
&& (t,q^i,q^i_t,q^i_{tt})\mapsto (t,q^i,q^i_t,\dot t=1,\dot
q^i=q^i_t,\dot q^i_t=q^i_{tt}),
\ee
where $J^2Q$, coordinated by $(q^\la,q^i_t,q^i_{tt})$, is the second order jet
manifold of the fiber bundle $Q\to\bR$. 
The affine bundle $J^2Q\to J^1Q$ is modelled over the vertical tangent bundle,
\beq
V_QJ^1Q\cong J^1Q\op\times_QVQ, \label{gm217}
\eeq
of the affine jet bundle $J^1Q\to Q$.

The jet manifold $J^1Q$ is provided with the canonical tangent-valued form,
\be
\wh v= \th^i\ot \dr^t_i, 
\ee
where $\th^i=dq^i-q^i_tdt$ are the contact forms. We have the corresponding 
endomorphism, 
\be
\wh v(\dr_t) = -q^i_t\dr_i^t,  \qquad \wh v(\dr_i)=\dr^t_i, \qquad
\wh v(\dr^t_i)=0,
\ee
of the tangent bundle $TJ^1Q$ and that,
\be
\wh v(dt)=0,\qquad \wh v(dq^i)=0, \qquad 
\wh v(dq^i_t) =\th^i,
\ee
of the cotangent bundle $T^*J^1Q$ of $J^1Q$. The nilpotent rule $\wh v^2=0$
holds. 

Due to the imbeddings (\ref{z260}),  any connection,
\be
\G=dt\ot (\dr_t +\G^i\dr_i), 
\ee
on a fiber bundle $Q\to\bR$  can be identified with a nowhere vanishing 
horizontal vector field,
\beq
\G = \dr_t + \G^i \dr_i, \label{a1.10}
\eeq
on $Q$ which is the horizontal lift of the
standard vector field $\dr_t$ on $\bR$ by means of $\G$. Conversely, any
vector field $\G$ on $Q$ such that
$dt\rfloor\G =1$ defines a connection on $Q\to\bR$. Accordingly,  the
covariant differential,
\be
 D_\G: J^1Q\op\to_Q VQ, \qquad  \dot q^i\circ D_\G
=q^i_t-\G^i,
\ee
associated with a connection $\G$ on
$Q\to\bR$, takes its values into the vertical tangent bundle $VQ$ of
$Q\to\bR$.

\begin{rem} From the
physical viewpoint, a connection (\ref{a1.10}) sets a reference frame.
There is one-to-one correspondence between these connections  and the
equivalence classes of atlases of local constant trivializations of the fiber
bundle $Q\to\bR$, i.e., such that transition functions
$q^i\to q'^i$ of the corresponding bundle coordinates are independent of $t$,
and
$\G=\dr_t$ with respect to these coordinates.$^{13-15}$ In particular, every
trivialization of $Q$ defines a complete connection $\G$ on $Q\to \bR$,
and {\it vice versa}. 
\end{rem}

A connection $\xi$ on the jet bundle $J^1Q\to \bR$ is said to be holonomic
if it is a section, 
\ben
&& \xi=\dr_t + q^i_t\dr_i + \xi^i \dr_i^t, \label{a1.30} \\
&& dt\rfloor\xi=1, \qquad \xi\rfloor\wh v=0, \nonumber
\een
of the holonomic sub-bundle
$J^2Q\to J^1Q$ of the affine jet bundle $J^1J^1Q\to J^1Q$. 
Holonomic connections (\ref{a1.30}) make up an affine space modelled over the
linear space of vertical vector fields on the affine jet bundle $J^1Q\to Q$,
i.e., which live in $V_QJ^1Q$. 
Every holonomic connection $\xi$ defines the corresponding covariant
differential on the jet manifold $J^1Q$:
\ben
&& D_\xi: J^2Q\ar_{J^1Q} V_QJ^1Q\subset VJ^1Q, \nonumber\\
&& \dot q^i \circ D_\xi =0, \qquad \dot q^i_t\circ D_\xi= q^i_{tt} - \xi^i,
\label{gm271}
\een
which
takes its values into the vertical tangent bundle $V_QJ^1Q$ of the affine jet
bundle $J^1Q\to Q$. Any
integral section
$\ol c:\bR\supset ()\to J^1Q$ for a holonomic connection
$\xi$ is holonomic, i.e., $\ol c=\dot c$ where $c$ is a curve in $Q$.

A second order dynamic equation (or simply a dynamic equation) on a
configuration bundle $Q\to\bR$ is defined as the
kernel, 
\beq
q^i_{tt}=\xi^i(t,q^j,q^j_t), \label{z273}
\eeq
of the covariant differential (\ref{gm271}) for some holonomic connection
$\xi$ on the jet bundle $J^1Q\to\bR$. Therefore, holonomic connections are
also called dynamic equations. By a solution of the dynamic equation
(\ref{z273}) is meant a curve $c$ in $Q$ whose second order
jet prolongation
$\ddot c$ lives in (\ref{z273}). Any integral section $\ol
c$ for the holonomic connection $\xi$ is the jet prolongation $\dot c$ of a
solution
$c$ of the dynamic equation (\ref{z273}), 
and {\it vice versa}.
\bigskip 

\noindent 
{\bf III. NONHOLONOMIC CONSTRAINTS}
\bigskip

Let $\bS$ be an $n$-dimensional codistribution 
on the velocity phase
space $J^1Q$.
Its annihilator
$\rA(\bS)$ is treated as a nonholonomic constraint.
Let the codistribution $\bS$ be locally
spanned by the 1-forms,
\be
s^a= s^a_0dt + s^a_idq^i +\dot s^a_idq^i_t, 
\ee
on the jet manifold $J^1Q$. Then a dynamic equation $\wt\xi$ on the
configuration bundle $Q\to\bR$ is said to be compatible with the
nonholonomic constraint
$\bS$ if
\be
s^a(\wt\xi)=\wt\xi\rfloor s^a=s^a_0 + s^a_iq^i_t +\dot s^a_i\wt\xi^i=0.
\ee
This equation is algebraically solvable for $n$ components of $\wt \xi$ iff 
the
$n\times m$ matrix $\dot s^a_i(q^\la,q^i_t)$ has everywhere 
maximal rank $n\leq m$. Therefore, we
restrict our consideration to the nonholonomic constraints, called admissible,
such that  
$\di \bS=\di \wh v(\bS)$.

If a nonholonomic constraint is admissible, there exists a local
$m\times n$ matrix $\dot s^i_a(q^\la,q^i_t)$ such that
\be
\dot s^i_a\dot s^b_i=\dl^b_a. 
\ee
Then the local decomposition (\ref{gm445}) of a dynamic equation $\xi$ can be
written in the form
\beq
\xi^i=\wt\xi^i +\dot s^i_as^a(\xi).
\label{gm450}
\eeq
The global decomposition (\ref{gm445}) exists by virtue of the following
lemma.

\begin{lem} The intersection 
\be
W=J^2Q\cap \rA(\bS)
\ee
 is an affine
bundle over $J^1Q$, modelled over the vector bundle 
\be
\ol W=V_QJ^1Q\cap\rA(\bS).
\ee
\end{lem}

\begin{proof}
$\ol W$ consists of the vertical vectors $v^i\dr_i^t\in V_QJ^1Q$
which fulfill the conditions 
\be
\dot s^a_i(q^\la,q^j_t)v^i=0. 
\ee
Since the nonholonomic constraint $\bS$ is admissible, every fiber of $\ol W$
is of dimension $m-n$, i.e., $\ol W$ is a vector bundle, while $W$ is an
affine bundle. 
\end{proof}

The affine structure of $W\to J^1Q$ implies that it has a global section
$\wt\xi$.

To construct the global decomposition (\ref{gm445}),
one usually perform a splitting of the vertical tangent bundle,
\beq
V_QJ^1Q=\ol W\op\oplus_{J^1Q} \cV, \label{gm452}
\eeq
and obtain the corresponding splitting of the  second order jet manifold,
\beq
J^2Q=W\op\oplus_{J^1Q} \cV. \label{gm453}
\eeq
Here $\cV\to J^1Q$ should be interpreted as the bundle of possible constraint
reaction accelerations.

 If an admissible nonholonomic constraint $\bS$ is of
dimension
$n=m$, a dynamic equation $\xi$  is decomposed in a unique fashion. If $n<m$,
the decomposition (\ref{gm445}) is not unique. Different variants of this
decomposition lead to different constraint reaction forces which, from the
physical viewpoint, characterize different types of nonholonomic constraints.
In next Section, we will construct the decomposition of dynamic equations of
Newtonian systems which corresponds to ideal nonholonomic constraints.

Now, let us consider some important examples of nonholonomic
constraints.

Let $N$ be a closed imbedded submanifold of the velocity phase space $J^1Q$,
defined locally by  the equations
\be
f^a(q^\la,q^i_t)=0, \qquad a=1,\ldots, n<m. 
\ee
One can treat $N$ as a nonholonomic constraint given by the
codistribution $\bS=\rA(TN)$ on $J^1Q\mid_N$. This codistribution is locally
spanned by the 1-forms
\be
s^a= df^a=\dr_tf^adt +\dr_jf^adq^j +\dr_j^tf^adq^j_t. 
\ee
The nonholonomic constraint $N$ is admissible iff the matrix
$(\dr_j^tf^a)$ is of maximal rank $n$. It follows that $N$ is a fibred
submanifold of the affine jet bundle $J^1Q\to Q$.

A nonholonomic constraint $N$ is said to be  linear if it is an affine
sub-bundle of the affine jet bundle $J^1Q\to Q$. Locally, a linear
constraint $N$ is given by the
 equations
\beq
f^a=f^a_0(q^\la) + f^a_i(q^\la)q^i_t=0, \label{gm520}
\eeq
where the matrix $f^a_i$ is of maximal rank. A linear constraint 
is always admissible. Since $N$ is an affine sub-bundle of $J^1Q\to Q$, it has
a global section $\G$ which is a connection on the configuration bundle
$Q\to\bR$, called the constraint reference frame.  Then, the
constraint equations (\ref{gm520}) take the form
\beq
f^a_i(q^\la)(q^i_t-\G^i)=0. \label{gm521}
\eeq
We can say that the linear constraint is immovable with respect to the
constraint reference frame $\G$. Then, one can think of $\dot
q^i_\G=q^i_t-\G^i$, satisfying the equation (\ref{gm521}), as  virtual
 velocities relative to the linear  constraint
$N$. 

Let now a configuration space $Q$ admit a composite fibration $Q\to \Si\to 
\bR$, where
\be
\pi_{Q\Si}:Q\to \Si
\ee
 is a fiber bundle, and let $(t,\si^r,q^a)$  be
coordinates on
$Q$, compatible with this fibration. Given a connection,
\beq
B =dt\ot(\dr_t +B^a\dr_a) + d\si^r\ot(\dr_r +B^a_r\dr_a), \label{gm461}
\eeq  
on the fiber bundle $Q\to \Si$, we have the corresponding
horizontal splitting of the tangent bundle
$TQ$. Restricted to the jet manifold $J^1Q\subset TQ$, this splitting reads
\be
&& J^1Q=B(\pi_{Q\Si}^*J^1\Si)\op\oplus_QV_\Si Q, \\
&& \dr_t +\si^r_t\dr_r + q^a_t\dr_a= [(\dr_t + B^a\dr_a) + \si^r_t
(\dr_r+ B^a_r\dr_a)] + [ q^a_t-B^a-\si^r_tB^a_r]\dr_a, 
\ee
where $\pi_{Q\Si}^*J^1\Si$ is the pull-back of the affine jet
bundle $J^1\Si\to\Si$ onto
$Q$. It is readily observed that 
\be
N=B(\pi_{Q\Si}^*J^1\Si)
\ee 
is an affine sub-bundle of the affine jet bundle 
$J^1Q\to Q$, defined locally by the equations
\be
q^a_t-\si^r_tB^a_r(q^\la)-B^a(q^\la)=0. 
\ee
This sub-bundle yields a linear nonholonomic
constraint.$^{16,17}$ The
corresponding  codistribution  $\bS=\rA(TN)$ is locally spanned by 
the 1-forms,
\ben
&& s^a=-(\dr_tB^a +\si^r_t\dr_tB^a_r)dt -(\dr_sB^a +\si^r_t\dr_sB^a_r)d\si^s 
- \label{gm464}\\
&& \qquad (\dr_bB^a +\si^r_t\dr_bB^a_r)dq^b + dq^a_t 
- B^a_rd\si^r_t. \nonumber
\een
With the connection (\ref{gm461}), we also have the 
splitting of the vertical tangent bundle $VQ$ of $Q\to\bR$
and the corresponding splitting of the
vertical tangent bundle $V_QJ^1Q$ which reads
\ben
&& V_QJ^1Q=\ol W\op\oplus_{J^1Q} \cV,\nonumber\\
&& \dot\si^r_t\dr_r^t + \dot q^a_t\dr_a^t= \dot\si^r_t(\dr^t_r +B^a_r\dr^t_a)
+(\dot q^a_t -B^a_r\dot\si^r_t)\dr_a^t. \label{gm465}
\een
It is readily observed that $\ol W\mid_N$ consists of
vertical vectors which are the annihilators of the
codistribution (\ref{gm464}). The splitting (\ref{gm465}) yields the
corresponding splitting (\ref{gm453}) of the second order jet manifold $J^2Q$.
Then we obtain the decomposition  (\ref{gm445}) of every dynamic equation
$\xi$ on
$J^1Q$ as
\be
\wt\xi^r=\xi^r, \qquad \wt\xi^a=\xi^a-s^a(\xi).
\ee

\bigskip 

\noindent 
{\bf IV. NEWTONIAN SYSTEMS WITH NONHOLONOMIC CONSTRAINTS}
\bigskip

Let $Q\to\bR$ be a fiber bundle
together with (i)
 a non-degenerate
fiber metric, 
\be
 \wh m:J^1Q\to V^*Q\op\ot_Q V^*Q,\qquad \wh m=\frac12 m_{ij} \ol dq^i\vee \ol
dq^j, 
\ee
 in the fiber bundle $V_QJ^1Q\to J^1Q$ which
satisfies the symmetry condition,
\beq
\dr_k^t m_{ij}=\dr_j^t m_{ik}, \label{a1.103} 
\eeq
 and (ii) a dynamic equation $\xi$ (\ref{a1.30}) on the
jet bundle $J^1Q\to \bR$, related to the fiber metric $\wh m$ by the
compatibility condition, 
\beq
2\xi\rfloor dm_{ij} + m_{ik}\dr_j^t\xi^k + m_{jk}\dr_i^t\xi^k = 0.
\label{a1.95}
\eeq
The triple $(Q,\wh m,\xi)$ is called a Newtonian system.$^{15}$ A Newtonian
system is said to be standard if $\wh m$ is the pull-back of a fiber 
metric in
the vertical tangent bundle $VQ$ in accordance with the isomorphism
(\ref{gm217}). In this case, $\wh m$ is independent of the velocity
coordinates $q^i_t$.

The notion of a Newtonian system generalizes the second Newton law of particle
mechanics. Indeed, the dynamic equation for a
Newtonian system is  equivalent to the equation   
\beq
m_{ik}(q^k_{tt}-\xi^k)=0. \label{z355}
\eeq
Therefore, $\wh m$ is called a mass metric.

There are two main reasons in order to consider Newtonian systems. From the
physical viewpoint, with   a mass metric, we can introduce the notion of an
external force, defined as a section of the vertical cotangent bundle
$V_Q^*J^1Q\to J^1Q$.  Let $(Q,\wh m,\xi)$ be a Newtonian system and $F$ an
external force. Then
\be
\xi^i_F= \xi^i + (m^{-1})^{ik}F_k,
\ee
is a dynamic equation, but the triple $(Q,\wh m,\xi_F)$ is  a Newtonian
system only if $F$
possesses the property
\beq
\dr_i^tF_j+\dr_j^tF_i=0.  \label{gm383}
\eeq From the mathematical viewpoint, 
the equation (\ref{z355}) is the kernel of an Euler--Lagrange-type operator. 
By an appropriate choice of a mass metric, one may hope to bring it into 
Lagrange equations. This is the well-known inverse problem in time-dependent
mechanics.

Here, we consider  Newtonian systems because they provide the vertical 
tangent bundle
$V_QJ^1Q$ with a nondegenerate fiber metric $\wh m$. Let us assume that $\wh
m$ is a Riemannian metric.  With this metric, we immediately obtain
 the splitting (\ref{gm452}), where
$\cV$ is the orthocomplement of
$\ol W$. Then the corresponding decomposition (\ref{gm450}) takes 
the form,$^9$
\beq
\xi^i=\wt\xi^i + \wt m_{ab}m^{ij}\dot s^a_js^b(\xi), \label{gm454}
\eeq
where $\wt m_{ab}$ is the inverse matrix of 
\be
\wt m^{ab}= \dot s^a_i\dot s^b_jm^{ij}.
\ee

It is readily observed that the decomposition (\ref{gm454}) satisfies the
generalized D'Alembert principle. The constraint reaction acceleration, 
\beq
-r^i= - \wt m_{ab}m^{ij}\dot s^a_js^b(\xi), \label{jmp5}
\eeq
is orthogonal to every element of $V_QJ^1Q\cap \rA(\bS)$ with respect to the
mass metric $\wh m$. Since elements of  $V_QJ^1Q\cap \rA(\bS)$ can be
treated as the virtual accelerations relative to the 
nonholonomic constraint $\bS$, the constraint reaction acceleration
(\ref{jmp5}) characterizes $\bS$ as an ideal constraint. 

The  Gauss principle is also fulfilled as
follows.
Given a dynamic equation $\xi$ and the above-mentioned fiber metric $\wh m$,
let us define a positive function $G(w)$ on $J^2Q$ as
\be
&& G(w)=\wh m\left(\xi(\pi^2_1(w))- w,\xi(\pi^2_1(w))- w\right),\\ 
&& G(q^\la,q^i_t,q^i_{tt})=m_{ij}(q^\la,q^k_t)(\xi^i(q^\la,q^k_t)- q^i_{tt})
(\xi^j(q^\la,q^k_t)- q^j_{tt}). 
\ee
We say that $\| w\|=G(w)^{1/2}$
is a norm of $w\in J^2Q$.

\begin{prop} Among all dynamic equations compatible with a
nonholonomic constraint, the dynamic equation $\wt\xi$ defined by the
decomposition (\ref{gm454}) is that of least norm.
\end{prop}

\begin{proof}
Let $\zeta$ be another dynamic equation which takes its values into $W$. Then 
$\wt\xi-\zeta\subset\ol W$ and 
\be
\wh m(\wt\xi-\zeta, \xi-\wt\xi)=0.
\ee
Hence, we obtain
\be
\|\zeta\|=\wh m(\xi-\wt \xi +\wt\xi-\zeta,\xi-\wt \xi +\wt\xi-\zeta)=
\|\wt\xi\| + \wh m(\wt\xi-\zeta,\wt\xi-\zeta).
\ee
\end{proof}

In next Section, we will show that, in the case of nondegenerate Lagrangian
systems and linear nonholonomic constraints, the decomposition (\ref{gm454})
satisfies the traditional D'Alembert principle.

\bigskip 

\noindent 
{\bf IV. LAGRANGIAN SYSTEMS WITH NONHOLONOMIC CONSTRAINTS}
\bigskip

Nondegenerate Lagrangian systems are particular Newtonian systems.

A Lagrangian is defined as a horizontal density, 
\beq
 L= \cL dt,  \qquad \cL: J^1Q\to\bR, \label{a1.81}
\eeq
on the velocity phase space $J^1Q$. Here, we apply in a 
straightforward manner
the first variational formula.$^{13,15}$

Let us consider a projectable vector field
\be
u=u^t\dr_t +u^i\dr_i, \qquad u^t=0,1, 
\ee
on the configuration bundle $Q\to\bR$ and calculate the Lie derivative of
the Lagrangian  (\ref{a1.81}) along the jet prolongation, 
\be
\ol u=u^t\dr_t +u^i\dr_i +d_tu^i\dr_i^t, 
\ee
 of $u$, where $ d_t=\dr_t+q^i_t\dr_i +\cdots$ is the
operator of formal derivative. We obtain
\beq
\bL_{\ol u}L =(\ol u\rfloor d\cL) dt=
(u^t\dr_t +u^i\dr_i +d_tu^i\dr_i^t)\cL dt. \label{1004}\\
\eeq

The first variational formula provides the following canonical
decomposition of the Lie derivative (\ref{1004}) in accordance with the
variational problem:
\beq
\ol u\rfloor d\cL=
(u^i-u^tq^i_t)\cE_i +d_t(u\rfloor H_L) \label{m218}
\eeq
where
\beq
H_L= \wh v(dL) + L=\pi_i dq^i -(\pi_iq^i_t-\cL)dt \label{z331}
\eeq
is the Poincar\'e--Cartan form and
\ben
&& \cE_L: J^2Q\to V^*Q,\nonumber\\
&&\cE_L=\cE_i\th^i=(\dr_i-d_t\dr^t_i)\cL \th^i,\label{983}
\een
is the Euler--Lagrange operator
for $L$.  We will use the notation 
\be
\pi_i=\dr^t_i\cL, \qquad \pi_{ji}=\dr_j^t\dr_i^t\cL.
\ee
A Lagrangian $L$ is called  nondegenerate if 
${\rm det}\,\pi_{ji}\neq 0$
everywhere on the velocity phase space $J^1Q$.

The kernel $\Ker\cE_L\subset J^2Q$ of the Euler--Lagrange operator
(\ref{983})  defines the system of second order differential equations, 
\beq
(\dr_i-d_t \dr^t_i)\cL=0, \label{b327}
\eeq
on $Q$, called the Lagrange equations. 
Their solutions are (local) section $c$ of the fiber bundle $Q\to\bR$
whose second order jet prolongations $\ddot c$ live in (\ref{b327}).

A holonomic connection on the jet bundle $J^1Q\to\bR$ is said to be a
Lagrangian connection $\xi_L$ for the Lagrangian $L$ if it takes its values in
the kernel (\ref{b327}) of the Euler--Lagrange operator $\cE_L$. Every
Lagrangian connection $\xi_L$ defines a dynamic equation on the
configuration space $Q$ whose solutions are also solutions of the Lagrange
equations (\ref{b327}).  If $L$ is
non-degenerate, the Lagrange equation (\ref{b327}) can be algebraically
solved  for the second order derivatives, and they are equivalent to the
dynamic equation,
 \beq
q^i_{tt}=\xi_L^i, \qquad \xi_L^i=(\pi^{-1})^{ij}\cE_j+ q^i_{tt},  
\label{a1.91}
\eeq
called the Lagrange dynamic equation.

Every Lagrangian $L$ on the jet manifold $J^1Q$ yields the 
 Legendre map,
\beq
\wh L:J^1Q\to V^*Q,\qquad p_i \circ\wh L = \pi_i, \label{a303}
\eeq
where $(t,q^i,p_i)$ are holonomic coordinates on the vertical cotangent
bundle $V^*Q$. 
As is well known, the Legendre map (\ref{a303}) is a local diffeomorphism iff
$L$ is nondegenerate.  A Lagrangian $L$ is called  hyperregular if the
Legendre map $\wh L$ is a diffeomorphism.

The vertical tangent map $V\wh L$ to the Legendre map $\wh
L$ reads
\be
V\wh L: V_QJ^1Q\to VV^*Q\cong V^*Q\op\times_QV^*Q.
\ee
It yields the linear fibred morphism
$V_QJ^1Q\to V_Q^*J^1Q$ and the corresponding mapping,
\beq
J^1Q\to V_Q^*J^1Q\op\ot_{J^1Q}V_Q^*J^1Q, \qquad m_{ij}=\pi_{ij}. 
\label{gm380}
\eeq
If a Lagrangian $L$ is nondegenerate, then (\ref{gm380}) is 
a mass metric which
satisfies the symmetry condition (\ref{a1.103}) and the compatibility
condition  (\ref{a1.95}) for the Lagrange dynamic equation (\ref{a1.91}). 

Thus, every
nondegenerate Lagrangian $L$ defines a Newtonian system. Moreover, a
nondegenerate Lagrangian system plus an external force which fulfills the
condition (\ref{gm383}) is also a Newtonian system. Conversely, every
standard Newtonian system can be seen as a Lagrangian system with the
Lagrangian,
\beq
L=\frac12m_{ij}(q^i_t-\G^i)(q^j_t- \G^j)dt, \label{jmp7}
\eeq
where $\G$ is a reference frame, plus an external force.

Given a nondegenerate Lagrangian $L$ with a Riemannian mass metric
$m_{ij}=\pi_{ij}$, 
let now $\bS$ be an admissible nonholonomic constraint on
the velocity phase space
$J^1Q$. Since this is a particular Newtonian system, we obtain  the dynamic
equation 
\ben
&& q^i_{tt} = \xi_L^i - \wt m_{ab}m^{ij}\dot s^a_j(\dot s^b_k\xi^k_L
+s^b_k q^k_t + s^b_0), \label{gm456}\\
&& \xi^i_L  = m^{ij}(-\dr_t\pi_j - \dr_k\pi_jq^k_t +\dr_j\cL),
\nonumber
\een
which is compatible with the constraint
$\bS$, treated as an ideal nonholonomic constraint.
This is the Lagrange dynamic equation in the presence of the additional
constraint reaction force
\beq
F_i=-\wt m_{ab}\dot s^a_is^b(\xi_L). \label{gm522}
\eeq
Let us consider the energy conservation law in the presence of this
force.

The energy conservation law in Lagrangian time-dependent mechanics is 
deduced from the first
variational formula (\ref{m218}) when the vector field $u=\G$ is a 
reference
frame. On the shell $\cE_i=0$ (\ref{b327}), this formula leads to the weak
identity,
\beq
\bL_{\ol\G}L \ap- d_t(\pi_i\dot q^i_\G -\cL),
\label{m227}
\eeq
where $\dot q^i_\G=q^i_t -\G^i$ is a relative velocity and
\beq
T_\G= \pi_i\dot q^i_\G -\cL \label{m228}
\eeq
is the energy function with respect to
the reference frame $\G$.$^{14,15,18}$  In the presence of an
external force
$F$, i.e., on the shell $\cE_i=-F_i$, the weak identity (\ref{m227}) is
modified as
\be
\bL_{\ol\G}L-\dot q^i_\G F_i=-d_t T_\G. 
\ee
 It is readily observed that, if a
nonholonomic constraint is linear and
$\G$ is a constraint reference frame, the constraint
reaction force (\ref{gm522}) does not contribute to the energy conservation
law. It follows that, in this case, the standard D'Alembert principle holds,
while the equation (\ref{gm456}) describes a motion in the 
presence of an ideal
nonholonomic constraint in the spirit of this principle. 

The constrained  equation of motion (\ref{gm456}) 
is neither Lagrange equations nor a dynamic equation of 
a Newtonian system. In
Section VI, we aim to show that it can be seen as a part of Hamilton equations
in the framework of the Hamiltonian formalism extended to the configuration
space $VQ$.
\bigskip 

\noindent 
{\bf V. VERTICAL EXTENSION OF HAMILTONIAN FORMALISM}
\bigskip

This Section provides a brief exposition of Hamiltonian formalism of
time-dependent mechanics on a configuration bundle $Q\to\bR$ and its
extension to the vertical configuration space $VQ$. We consider this
extension because any first order dynamic equation on the 
momentum phase space
$V^*Q$ can be seen as a Hamilton equation in the framework of the extended
Hamiltonian formalism. This extension is also of interest in the
path-integral formulation of mechanics.$^{19,20}$

Given a mechanical system on a configuration bundle $Q\to\bR$, its
momentum phase space is the vertical cotangent bundle $V^*Q$ of $Q\to\bR$,
equipped with the holonomic coordinates $(t,q^i,p_i=\dot q_i)$.$^{13-15}$
The momentum phase space $V^*Q$ is endowed with the canonical exterior
3-form,
\be
\bom=dp_i\w dq^i\w dt. 
\ee

Let us consider the cotangent
bundle
$T^*Q$ of $Q$ with the holonomic coordinates $(t,q^i,p,p_i)$. It admits the
canonical Liouville form  
\beq
\Xi=pdt +p_idq^i. \label{m91}
\eeq
 An exterior 1-form $H$  on the momentum phase space $V^*Q$ is called a 
Hamiltonian form if it is the pull-back
\beq
H=h^*\Xi= p_i dq^i -\cH dt \label{b4210}
\eeq
of the Liouville form $\Xi$ (\ref{m91}) 
by a section $h$ of the fiber bundle
\beq
\zeta:T^*Q\to V^*Q. \label{y1}
\eeq

\begin{rem}
With respect to a trivialization $Q\cong \bR\times M$, the 
Hamiltonian form (\ref{b4210}) is the well-known  integral invariant of
Poincar\'e--Cartan, where $\cH$ is a Hamiltonian. The peculiarity of
Hamiltonian time-dependent mechanics issues from the fact that Hamiltonians
are not scalar functions under time-dependent transformations, but make up an
affine space modelled over the linear space of functions on $V^*Q$.
\end{rem}

For instance, every connection $\G$ on a configuration bundle $Q\to \bR$ 
is an affine section,
\be
p\circ\G=-p_i\G^i,
\ee
of the fiber bundle (\ref{y1}), and defines the
Hamiltonian form 
\be
H_\G=p_idq^i -p_i\G^idt.
\ee
It follows that any Hamiltonian form on the momentum 
phase space $V^*Q$ admits
the splitting,  
\be
H =H_\G-\wt{\cH}_\G dt =p_idq^i -(p_i\G^i +\wt{\cH}_\G) dt,  
\ee
where $\G$ is a connection on $Q\to\bR$ and $\wt\cH_\G$ is a real
function on $V^*Q$, called the  Hamiltonian 
function.
The following assertions are basic facts in the Hamiltonian formulation of
time-dependent mechanics.$^{14,15}$

\begin{prop}\
Every Hamiltonian form $H$ on the momentum phase space
$V^*Q$ defines the associated Hamiltonian map, 
\be
\wh H:V^*Q\to J^1Q,\qquad q_t^i\circ\wh H=\dr^i\cH. 
\ee
\end{prop}

\begin{prop} Given a Hamiltonian form $H$ on the momentum phase space
$V^*Q$, there exists a unique connection 
\beq
\g_H=\dr_t +\dr^i\cH\dr_i -\dr_i\cH\dr^i \label{m57}
\eeq
 on 
$V^*Q\to \bR$, called a Hamiltonian connection, such that
\be
\g_H\rfloor\bom= dH. 
\ee
\end{prop}

The kernel of the covariant differential of the Hamiltonian connection
(\ref{m57}) defines the Hamilton equations,
\bea
&& q^i_t =\dr^i\cH, \label{m41a}\\
&&  p_{ti} =-\dr_i\cH, \label{m41b}
\eea
for the Hamiltonian form $H$. Their solutions are integral curves for the
Hamiltonian connection $\g_H$ (\ref{m57}).

Now let us consider the vertical tangent bundle  
$VQ$ of the fiber bundle
$Q\to\bR$, coordinated by
$(t,q^i,\dot q^i)$.   It can be seen as a new configuration space, called
the vertical configuration space. 
The corresponding vertical momentum phase
space is the vertical cotangent bundle 
$V^*VQ$ of  
$VQ\to \bR$. 
The vertical momentum phase space $V^*VQ$ is canonically isomorphic to the
vertical tangent bundle 
$VV^*Q$  of the ordinary momentum phase space
$V^*Q\to\bR$, coordinated by  $(t, q^i, p_i,\dot q^i,\dot p_i)$.
It is easily seen from the transformation laws
that $(q^i, \dot p_i)$ and $(\dot q^i, p_i)$ are canonically conjugate
pairs.

The vertical momentum phase space $VV^*Q$ is endowed with the 
canonical 3-form,
\be
\bom_V=[d\dot p_i\w dq^i +dp_i\w d\dot q^i]\w dt.
\ee
For the sake of brevity, one can write $\bom_V=\dr_V\bom$,
where $\dr_V=\dot q^i\dr_i +\dot p_i\dr^i$ is the vertical derivative.

The notions of a Hamiltonian connection, a
Hamiltonian form, etc., on the vertical momentum phase space 
$VV^*Q\cong V^*VQ$ are introduced similarly to those on the ordinary momentum
phase space $V^*Q$. In particular, a Hamiltonian form
on $VV^*Q$ reads
\be
H_V=\dot p_idq^i + p_i d\dot q^i-\cH_Vdt.
\ee
Since Hamiltonian forms are determined modulo exact forms and the function 
$p_i\dot q^i$ is globally defined on $VV^*Q$, we will write 
\beq
H_V=\dot p_idq^i -\dot q^idp_i -\cH_Vdt. \label{m148}
\eeq
The corresponding Hamilton equations read
\bea
&&\g^i= q^i_t=\dot \dr^i\cH_V, \label{z700a}\\
&& \g_i=p_{ti}=-\dot \dr_i\cH_V, \label{z700b}\\
&& \ol \g^i=\dot q^i_t=\dr^i\cH_V, \label{z700c}\\
&& \ol\g_i=\dot p_{ti}= -\dr_i\cH_V, \label{z700d}
\eea
where
\be
\ol\g =\dr_t +\g^i\dr_i +\g_i\dr^i +\ol \g^i\dot\dr_i+\ol \g_i\dot\dr^i
\ee
is a Hamiltonian connection on the vertical momentum phase space $VV^*Q\to
\bR$.

There is the following relation between Hamiltonian formalisms on $V^*Q$
and 
$VV^*Q$.$^{13,15}$ Let $VT^*Q$ be the vertical tangent bundle 
 of the cotangent bundle 
$T^*Q\to \bR$ is equipped with holonomic coordinates
$(t, q^i, p_i, p, \dot q^i, \dot p_i, \dot p)$ 
and endowed with the canonical form,
\be
\Xi_V=\dot p dt + \dot p_i dq^i - \dot q^i dp_i. 
\ee

\begin{prop}\label{p01} 
Let $\g_H$ be a Hamiltonian connection on the ordinary momentum phase space 
$V^*Q\to\bR$ for a Hamiltonian form,
\beq
H=h^*\Xi=p_id q^i-\cH dt. \label{z701}
\eeq
Then the connection 
\ben
&& V\g_H : VV^*Q\to VJ^1V^*Q\cong J^1VV^*Q, \nonumber \\
&& V\g_H = \dr_t
+\g^i\dr_i+\g^i\dr_i +\dr_V\g^i \dot\dr_i +\dr_V\g_i\dot\dr^i,
\label{43}
\een
on the vertical momentum phase space
$VV^*Q\to\bR$ is a Hamiltonian connection for the Hamiltonian form,
\ben
&& H_V=(Vh)^*\Xi_V=\dot p_idq^i -\dot q^idp_i -\dr_V\cH dt,  \label{m17}\\
&& \dr_V\cH=(\dot q^i\dr_i +\dot p_i\dr^i)\cH,
\een
where $Vh:VV^*Q\to VT^*Q$ is the vertical tangent map to
$h$.
\end{prop}

The corresponding Hamilton equations read
\bea
&& \g^i=\dot\dr^i\cH_V =\dr^i\cH,\label{z740a}\\
&& \g_i=-\dot\dr_i\cH_V =-\dr_i\cH,\label{z740b}\\
&& \ol \g^i=\dr^i\cH_V=\dr_V\dr^i\cH,\label{z740c}\\
&& \ol \g_i=-\dr_i\cH_V=-\dr_V\dr_i\cH.\label{z740d}
\eea
It is easily seen that the equations (\ref{z740a}) --
(\ref{z740b})  are exactly the Hamilton equations (\ref{m41a}) --
(\ref{m41b}) for the Hamiltonian form $H$. 

\begin{rem}
In order to clarify the physical meaning of the Hamilton equations
(\ref{z740c}) -- (\ref{z740d}), let $r(t)$ be a solution of the Hamilton
equations (\ref{z740a}) -- (\ref{z740b}).  Let 
$\dot r(t)$ be a Jacobi field, i.e., 
$r(t)+\ve \dot r(t)$
is also a solution of the same Hamilton equations modulo terms of order two in
$\ve$. Then it is readily observed that the Jacobi field $\dot r(t)$ fulfills
the Hamilton equations  (\ref{z740c}) -- (\ref{z740d}).
\end{rem}

The following assertion plays a prominent role in the sequel.$^{13,15}$

\begin{prop}\label{can4}  Any connection
$\g$ on the momentum phase space
$V^*Q\to\bR$ gives rise to the Hamiltonian connection, 
\beq
\g^i=\g^i, \quad \g_i=\g_i, \quad \ol\g^i =\dot p_j\dr^i \g^j-\dot
q^j\dr^i \g_j, \quad \ol\g_i = -\dot p_j\dr_i\g^j + \dot
q^j\dr_i\g_j, \label{m165}
\eeq
for the Hamiltonian form, 
\be
H_V=\dot p_i(dq^i-\g^idt) -\dot q^i(dp_i -\g_idt)  = 
\dot p_idq^i -\dot q^idp_i -(\dot p_i\g^i -\dot q^i\g_i)dt,
\ee
on the vertical momentum phase space $VV^*Q$. 
\end{prop}

In particular, if 
$\g$ is a  Hamiltonian connection on the fiber bundle
$V^*Q\to\bR$, then 
(\ref{m165}) is exactly the connection 
$V\g$ (\ref{43}).

It follows that every first order dynamic equation 
 on the momentum phase space
$V^*Q$ can be seen as the Hamilton equations
(\ref{z700a}) -- (\ref{z700b}) for a suitable 
Hamiltonian form on the vertical
momentum phase space.

\bigskip 

\noindent 
{\bf VI. HAMILTONIAN SYSTEMS WITH NONHOLONOMIC CONSTRAINTS}
\bigskip

Let $L$ be a hyperregular Lagrangian with a Riemannian mass metric $\wh m$.
In this case, Hamiltonian and Lagrangian
formalisms of time-dependent mechanics are equivalent. 
There exists a unique associated
Hamiltonian form $H$ (\ref{b4210}) on $V^*Q$ such that 
\bea
&& \wh H= \wh L^{-1}, \qquad p_i\equiv\pi_i(q^\la,\dr^j\cH(q^\la,p_k)), 
\qquad
q^i_t\equiv\dr^i\cH(q^\la,\pi_j(q^\la,q^k_t)), \label{m191a}\\
&& \cL\circ\wh H\equiv \g_H\rfloor H= p_i\dr^i\cH-\cH. \label{m191b}
\eea
As an immediate consequence of (\ref{m191a}), we have 
$J^1\wh H= (J^1\wh L)^{-1}$,
where 
the jet prolongations of the Hamiltonian  and  Legendre
maps read
\be
&& J^1\wh H: J^1V^*Q\to J^1J^1Q, \qquad
(q^\la,q^i_t, q^i_{(t)},q^i_{tt})\circ 
J^1\wh H=(q^\la,\dr^i\cH,q^i_t,d_t\dr^i\cH),\\
&& J^1\wh L:J^1J^1Q \to J^1V^*Q, \qquad (q^\la,p_i,q^i_t,p_{ti})
\circ J^1\wh L
=(q^\la,\pi_i, q^i_{(t)},d_t\pi_i).
\ee 
Then, using  (\ref{m191a}) -- (\ref{m191b}), we obtain 
\be
\g_H= J^1\wh L\circ \xi_L \circ\wh H.
\ee
Let introduce the notation $M^{ij}=\dr^i\dr^j\cH$. There are 
the relations
\be
M^{ik}(m_{kj}\circ\wh H)=\dl^i_j, \qquad 
m_{kj}(M^{ik}\circ\wh L) =\dl^i_j, \qquad m_{ij}=\pi_{ij}. 
\ee
It follows that $M$ is a fiber metric in the vertical tangent bundle
$V_QV^*Q$ of the fiber bundle $V^*Q\to Q$.

Given a codistribution $\bS$ on $J^1Q$, let us consider the
pull-back codistribution $\wh H^*\bS$ on $V^*Q$, 
spanned locally by 1-forms
\be
&&\bt^a=\wh H^*s^a=(s^a_0 +\dot s^a_j\dr_t\dr^j\cH)dt + (s^a_i +\dot
s^a_j\dr_i\dr^j\cH)dq^i + \dot s^a_iM^{ij}dp_j=  \\
&& \qquad \bt^a_0dt + \bt^a_idq^i +\dot \bt^{ai}dp_i. 
\ee
This codistribution defines a nonholonomic constraint 
on the momentum phase space $V^*Q$. 

Given a Hamiltonian connection $\g_H$ (\ref{m57}), let us find its splitting
\beq
\g_H=\wt \g + \vt \label{jmp3}
\eeq
where $\wt\g$ is a connection on $V^*Q\to\bR$ which satisfies the condition
\beq
\wt \g\subset \rA(\wh H^*\bS). \label{jmp4}
\eeq
The connection $\wt\g$ (\ref{jmp4}) obviously defines a first order dynamic
equation on the momentum phase space $V^*Q$ which is compatible with the
nonholonomic constraint $\wh H^*\bS$.
The
decomposition (\ref{jmp3}) is not unique. Let us construct it as follows.

Given a Hamiltonian connection $\g_H$, we consider the
codistribution
$S_H$ on
$V^*Q$, spanned locally by the 1-forms $dq^i-\g_H^idt$. Its annihilator
$\rA(S_H)$ is an affine sub-bundle of the affine jet bundle 
 $J^1V^*Q\to V^*Q$, modelled over
the vertical tangent bundle $V_QV^*Q$. The Hamiltonian connection $\g_H$
is a section of this sub-bundle. Let us take the intersection
\be
W=\rA(S_H)\cap \rA(\wh H^*\bS). 
\ee

\begin{lem}
$W$  is an affine bundle over $V^*Q$, modelled over
the vector bundle,
\be
\ol W=V_QV^*Q\cap \rA(\wh H^*\bS).
\ee
\end{lem}

\begin{proof} The intersection $\ol W$ consists of elements $v=v_i\dr^i$ of
$V_QV^*Q$ which fulfill the conditions
\be
v_i\dot \bt^{ai}=0.
\ee
 Since the nonholonomic constraint $\bS$ is 
admissible and the matrix $M^{ij}$ is nondegenerate, 
every fiber of $\ol W$ is
of dimension
$m-n$, i.e., $\ol W$ is a vector bundle, while $W$ is an affine bundle.
\end{proof}

Then, using the fiber metric $M$ in $V_QV^*Q$, we obtain the splitting
\be
V_QV^*Q = \ol W\oplus\cV,
\ee
where $\cV$ is the orthocomplement of $\ol W$, and the associated splitting 
\be
\rA(S_H)=W \oplus \cV.
\ee
 The corresponding decomposition (\ref{jmp3})
reads
\beq
\wt \g=\g_H - \wt M_{ab}M_{ij}\dot\bt^{ai}\bt^b(\g_H)\dr^j,\label{jmp6}
\eeq
where $\wt M_{ab}$ is the inverse matrix of 
\be
\wt M^{ab}=\dot\bt^{ai}\dot\bt^{bj}M_{ij}.
\ee

The splitting (\ref{jmp6}) is the Hamiltonian counterpart of the splitting
(\ref{gm454}). We have the relations 
\be
\wt m^{ab}= \wt M^{ab}\circ\wh H, \qquad \bt^a(\g_H)=s^a(\xi_L)\circ\wh H,
\ee
and as a consequence
\be
\wt\g=J^1L\circ\wt\xi\circ \wh H.
\ee

\begin{rem}
The above procedure can be extended in a straightforward manner to any
standard Newtonian system, seen as a Lagrangian system with the Lagrangian
(\ref{jmp7}) and an external force. Following this procedure, one may also
study a nonholonomic Hamiltonian system, without appealing to its Lagrangian
counterpart.
\end{rem}

The connection (\ref{jmp6}) defines the system of first 
order dynamic equations,
\beq
q^i_t=\dr^i\cH, \qquad p_{ti}=-\dr_i\cH -\wt
M_{ab}M_{ij}\dot\bt^{ai}\bt^b(\g_H),
\label{jmp8}
\eeq
on the momentum phase space $V^*Q$, which are not Hamilton equations.
Nevertheless, in accordance with Proposition \ref{can4}, one can restate the
constrained equations of motion (\ref{jmp8}) as the Hamilton equations
(\ref{z740a}) -- (\ref{z740b}) for the Hamiltonian form,
\be
H_V=\dot p_idq^i -\dot q^idp_i -\dr_V\cH dt -\dot q^i\wt
M_{ab}M_{ij}\dot\bt^{ai}\bt^b(\g_H)dt,
\ee
on the vertical momentum phase space $VV^*Q$, where the last term can be
written in brief as
$(-\dr_V\rfloor\vt\rfloor\bom)$. 

The Hamiltonian form of the constrained equations of motion may be
important in connection with the following speculations.

Given  a Hamiltonian form 
$H_V$
(\ref{m148}) on the vertical momentum phase space
$VV^*Q$, let us consider the
Lagrangian 
\beq
L_H= \dot p_iq^i_t -\dot q^ip_{ti} -\cH_V \label{z760}
\eeq
on the first order jet manifold $J^1VV^*Q$ of the fiber bundle
$VV^*Q\to\bR$. It is readily observed that the corresponding Lagrange
equations are exactly the Hamilton 
equations (\ref{z700a}) -- (\ref{z700d}) for
the  Hamiltonian form $H_V$. In particular, let 
$H$ be a Hamiltonian form on an ordinary momentum phase space 
$V^*Q$ and 
$\cH_V=\dr_V\cH$. In this case, the Lagrangian
(\ref{z760}) reads
\be
L_H= \dot p_i(q^i_t-\dr^i\cH) -\dot q^i(p_{ti} +\dr_i\cH). 
\ee
It is easily seen that this Lagrangian vanishes on solutions of the Hamilton
equations for the Hamiltonian form $H$. By this reason, it is applied to the
functional integral formulation of mechanics.$^{19,20}$

\end{document}